\documentclass[prl,aps,twocolumn,superscriptaddress,showpacs]{revtex4-1}

\usepackage{graphicx}
\usepackage{epstopdf}

\begin{document}

\title{From incommensurate correlations to  mesoscopic spin resonance in YbRh$_{2}$Si$_{2}$}

\author{C. Stock}
\affiliation{NIST Center for Neutron Research, 100 Bureau Drive, Gaithersburg, Maryland 20899, USA}
\affiliation{Indiana University, 2401 Milo B. Sampson Lane, Bloomington, Indiana 47404, USA}
\author{C. Broholm} 
\affiliation{Institute for Quantum Matter and Department of Physics and Astronomy, Johns Hopkins University, Baltimore, Maryland USA 21218}
\affiliation{NIST Center for Neutron Research, 100 Bureau Drive, Gaithersburg, Maryland 20899, USA}
\author{F. Demmel}
\affiliation{ISIS Facility, Rutherford Appleton Labs, Chilton, Didcot, OX11 0QX}
\author{J. Van Duijn}
\affiliation{Instituto de Investigaci—n en Energ'as Renovables and Departamento de F'sica Aplicada, Universidad de Castilla-La Mancha, 02006 Albacete, Spain}
\author{J. W. Taylor}
\affiliation{ISIS Facility, Rutherford Appleton Labs, Chilton, Didcot, OX11 0QX}
\author{H.J. Kang}
\affiliation{NIST Center for Neutron Research, 100 Bureau Drive, Gaithersburg, Maryland 20899, USA}
\author{R. Hu}
\affiliation{Condensed Matter Physics, Brookhaven National Laboratory, Upton, New York, USA 11973}
\author{C. Petrovic}
\affiliation{Condensed Matter Physics, Brookhaven National Laboratory, Upton, New York, USA 11973}

\date{\today}

\begin{abstract}

Spin fluctuations are reported near the magnetic field driven quantum critical point in YbRh$_{2}$Si$_{2}$.  On cooling, ferromagnetic fluctuations evolve into incommensurate correlations located at  ${\bf{q_{0}}}$=$\pm$ ($\delta$,$\delta$) with $\delta$=0.14 $\pm$ 0.04 r.l.u.  At low temperatures, an in plane magnetic field induces a sharp intra doublet resonant excitation at an energy  $E_{0}=g\mu_{B}\mu_{0}H$ with $g$=3.8$\pm$0.2.    The intensity is localized at the zone center indicating precession of spin density extending $\xi=6 \pm 2$ \AA\ beyond the $4f$ site.  

\end{abstract}

\pacs{PACS numbers: 75.40.Gb, 74.70.Tx, 75.50.Cc}
\maketitle

	The development of inter site coherence amongst Kondo-screened ions is a phenomenon that is central to understanding the heavy fermion state and quantum phase transitions.~\cite{HERTZ:1976ta,Gegenwart08:4,Lohneysen07:79,Ernst11:474}  Here we use neutron scattering to monitor such correlations as a function of magnetic field and temperature in a clean system where disorder due to chemical substitution is not expected.  These experiments show the formation of a coherent Kondo lattice and a low temperature field driven transition to a dressed single ion phase.  

	YbRh$_{2}$Si$_{2}$ is a heavy fermion metal with weak antiferromagnetic order at very low temperatures $\sim$ 50 mK.~\cite{Ishida03:68}  The resistivity varies linearly with temperature ($\rho \sim T$) demonstrating strong non-Fermi liquid character.~\cite{Trov00:85}  In a small magnetic field of $\sim$ 0.7 T along the $c$ axis (and $\sim$0.06 T when the field is applied within the $a-b$ plane), the magnetic order is suppressed and the non-Fermi liquid phase is transformed continuously into a Landau Fermi liquid where $\rho \sim T^{2}$.~\cite{Gegen02:89,Custers03:424}  The large change in Hall number  and band structure calculations indicate a quantum critical transition in the electron density which can be interpreted in terms of Fermi surface reconstruction or local quantum criticality.~\cite{Paschen04:432,Fried10:107,Norman05:71,Si01:413}  At higher fields de Haas van-Alphen measurements indicate a Lifshitz transition of the Fermi surface.~\cite{Rourke08:101} 
 
	Despite the heavy fermion nature, electron-spin resonance (ESR) shows a signal indicative of localized Yb$^{3+}$ $4f$ moment behavior under the application of a field.~\cite{Sichel03:91}  While some work has been performed on the magnetic dynamics,  little is known about the magnetic response across the field tuned quantum critical point.~\cite{Stockert06:378,Stockert07:1}   
	
	We present a neutron inelastic scattering study of single crystalline YbRh$_{2}$Si$_{2}$.  We will show that at zero field, the magnetic fluctuations are incommensurate.  Under the application of a magnetic field, driving the system through the quantum critical point, commensurate underdamped fluctuations develop at the Zeeman energy of the Yb$^{3+}$ $4f$ crystal field doublet.   The momentum localized resonance, however, indicates a spin density that extends beyond the $4f$ ion.
	
	Experiments on single crystalline YbRh$_{2}$Si$_{2}$ prepared using Zinc flux were performed on the SPINS and OSIRIS spectrometers and  powder experiments (discussed in the supplementary information) used MARI.  Approximately 300 (with total mass $\sim$ 3 g) single crystal samples (growth and characterization described in Ref. \onlinecite{Hu07:304}) were coaligned in the (HHL) scattering plane using a similar method as described previously.~\cite{Stock08:100}  Experiments on SPINS used a focussing analyzer with 11$^{\circ}$ acceptance,  E$_{f}$=3.7 meV and vertical magnetic fields up to 11 T.   To probe the low energy dynamics (E \textless 0.5 meV) we made use of the OSIRIS indirect spectrometer with E$_{f}$=1.84 meV.  The data on OSIRIS has been corrected for a background obtained by imposing detailed balance.~\cite{Hong06:74} In both experiments the temperature was monitored by a resistive thermometer attached near the sample.  The lowest temperature achieved was 100 mK, therefore this experiment does not access the ordered state.  Absolute normalization was performed against the (0,0,2) nuclear Bragg peak and through a comparison of the reported magnetic field dependent magnetization.~\cite{Gegen06:8}

\begin{figure}[t]
\includegraphics[width=8.95cm] {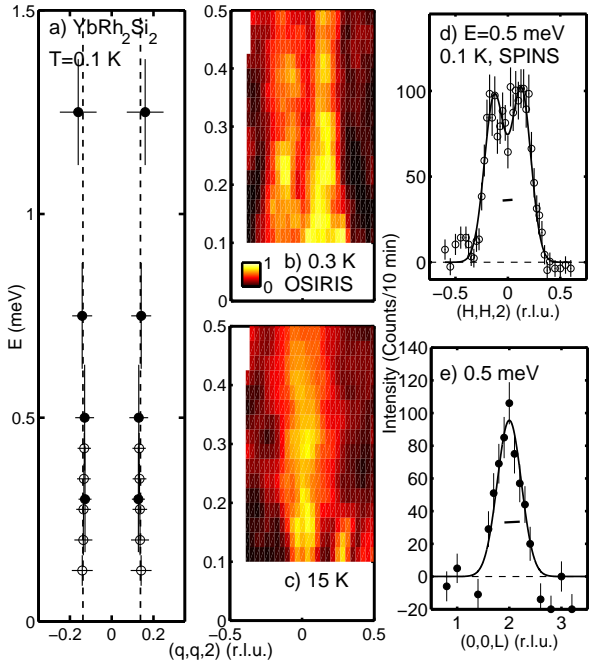}
\caption{\label{lowT_momentum} The $\hbar \omega-Q$ dependence of the magnetic fluctuations in YbRh$_{2}$Si$_{2}$ at T=0.1 K.  $(a)$ the position in $Q$ of the ridges of scattering where filled (open) circles are from SPINS (OSIRIS).  $(b)$ and $(c)$ plot contour maps of the spin fluctuations taken on OSIRIS at 0.3 K and 15 K integrating over the range (q,q,2$\pm$0.25). $(d)$ and $(e)$ show background corrected cuts in momentum at 0.5 meV.  The horizontal bars are the experimental resolution. }
\end{figure}

	We first describe the magnetic fluctuations at low temperatures near the ordering transition.  Figure \ref{lowT_momentum} shows constant energy scans near the ferromagnetic {\bf{Q}}=(0,0,2) position at 100 mK.  Panel $(a)$ summarizes the OSIRIS and SPINS data showing ridges extending along $\hbar \omega$.   This demonstrates that the momentum dependence is controlled by a higher energy scale such as the Kondo temperature or the Fermi energy.   As shown in the supplementary information, these incommensurate low-energy fluctuations are well below the first crystal field level located at 17.9 $\pm$ 0.6 meV and therefore are associated with allowed transitions within the ground state doublet.
	
	Panels $(d)$ and $(e)$ show representative scans taken on SPINS at 0.5 meV with a constant background subtracted.  The scan along the (H,H,0) direction ($d$) illustrates an incommensurate modulation in the basal plane while the scan along L (panel $(e)$) shows ferromagnetic inter plane correlations. An (H,H,2$\pm$0.25)-$\hbar \omega$ slice (panel $(b)$) demonstrates that the incommensurate ridges extend to the lowest energy transfers accessed ($\sim$ 0.1 meV).  The fluctuations are peaked at ${\bf{Q}}_{\perp}=(\delta, \delta)$ with $\delta$ =0.14 $\pm$ 0.4 with no measurable dispersion or offset along L.   We therefore expect ($\delta$, $\delta$) to be the characteristic wavevector of the low temperature spin density wave order corresponding to the transition reported using $\mu SR$ and susceptibility.~\cite{Ishida03:68,Trov00:85}  While there is no conclusive evidence that Fermi surface nesting drives this transition, the observed critical wavevector is not far from the spacing between large areas of the computed Fermi surface.~\cite{Norman05:71}   Low temperature diffraction and  calculations of $\chi({\bf{Q}})$ will be required for progress. 

	At elevated temperatures of (15 K, panel $(c)$), the in plane response changes considerably with the incommensurate fluctuations being replaced by commensurate ferromagnetic excitations forming a ridge at the zone center.  These results confirm  a competition between ferromagnetic $q$=0 excitations and incommensurate spin fluctuations as inferred from NMR based on a comparison between the Knight shift and relaxation rate.~\cite{Ishida02:89} 
	
\begin{figure}[t]
\includegraphics[width=9.3cm] {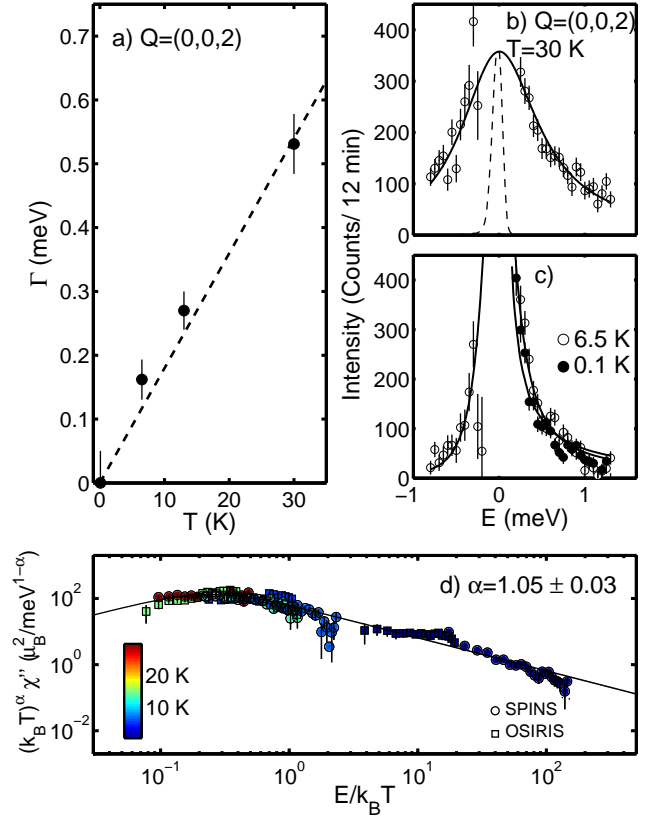}
\caption{\label{gamma_scale} $(a)$ illustrates the linewidth of the magnetic excitations taken from constant Q scans represented in panels $(b)$ and $(c)$.  The dashed curve in $(b)$ centered at E=0, is the measured resolution on SPINS.  Panel $(d)$ summarizes all of the temperature dependence from SPINS and OSIRIS and plots $k_{B}T ^{\alpha}\chi''$ as a function of $E/k_{B}T$ over the temperature range T=[0.1, 30] K (as described by the colorbar) and integrating  over {\bf{Q}}=$(\pm 0.1, \pm 0.1, 2 \pm0.25)$ on OSIRIS. The exponent $\alpha$ was fitted to be $\alpha$=1.05 $\pm$ 0.03 as described in the supplementary information.}
\end{figure}

	A more detailed survey of the $T$ dependent spin fluctuations can be found in Fig. \ref{gamma_scale}.  Constant {\bf{Q}}=(0,0,2) scans at 30 and 6.5 K are shown in panels $(b)$ and $(c)$.  The solid curves are a fit to the relaxational form $\chi''({\bf{Q}},\omega)=\chi'_{\bf{Q}}\omega\Gamma(T)/(\Gamma(T)^2+\omega^2)$, where $\Gamma(T)$ is the relaxation rate and $\chi'_{\bf{Q}}$ is the susceptibility.  The temperature dependence of $\Gamma(T)$ (Fig. \ref{gamma_scale} $(a)$) varies linearly and reaches zero close to T=0.  Such behavior is expected near quantum criticality.  

	In a similar manner to the case of scaling near a classical phase transition driven by thermal fluctuations, we investigate the scaling properties of the dynamic susceptibility near the low-field quantum critical point. Several scaling theories have been proposed for the susceptibility.~\cite{Millis93:48,Si01:413}  The uniform dynamic susceptibility $\chi''$ at all temperatures and energies from SPINS and OSIRIS are compiled in $(d)$ which shows $(k_{B}T)^{\alpha}\chi''$ as a function of $E/k_{B}T$.  The data is best fit with $\alpha$ =1.05 $\pm$ 0.03 consistent with 1 as expected in the case of an itinerant ferromagnetic at T$\textgreater$T$_{Curie}$ and $\Gamma \propto 1/\chi'_{\bf{q}}$, when $\chi'_{\bf{q}}$ has a Curie form.~\cite{Moriya79:14}  The exponent is the same as observed for the scaling of heat capacity.~\cite{Trov00:85}  Such a scenario is consistent with the fluctuations being close to ferromagnetic as reported in Fig. \ref{lowT_momentum} and are similar to the case of UCoGe which is close to a critical point between ferromagnetism and superconductivity.~\cite{Stock11:107} 

 \begin{figure}[t]
\includegraphics[width=9.3cm] {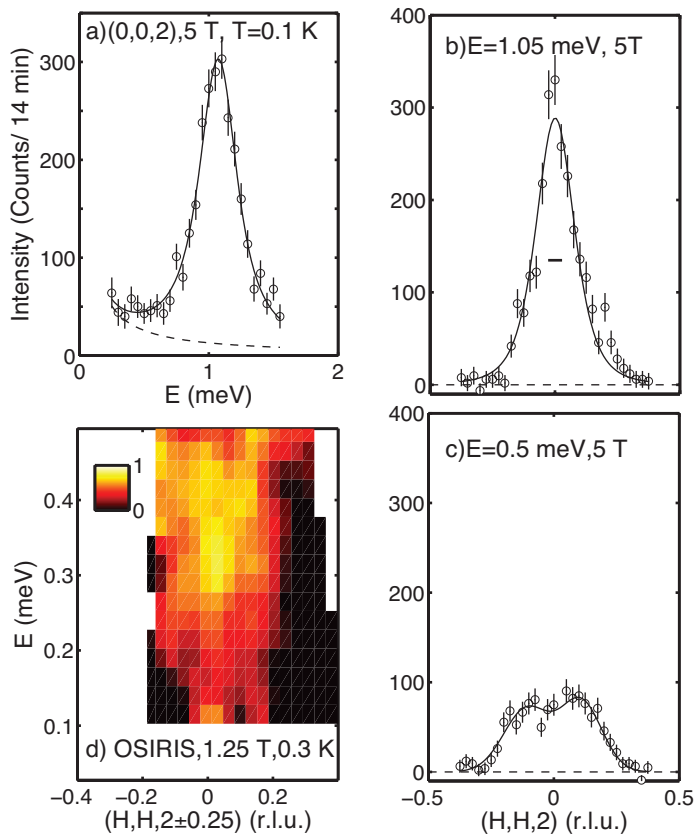}
\caption{\label{resonance_figure} $(a)$ illustrates a background subtracted constant Q scan showing the presence of a sharp resonance peak on the application of a 5T field on SPINS.  $(b)$ and $(c)$ illustrate constant energy scans at the resonance energy and below at 0.5 meV.  $(d)$ illustrates a contour taken on OSIRIS illustrating that the resonance peak at 1.25 T is sharp in H and energy.}
\end{figure}

We now investigate the possibility of anomalous scaling in our experiment which would manifest as a deviation from the conventional scaling described above.~\cite{Millis93:48}  A fit to $k_{B}T^{\beta}\chi''$ as a function of $E/(k_{B}T)^\beta$ resulted in $\beta$=0.92 $\pm$0.05  also consistent with 1 and different from other scenarios used to describe the fluctuations in CeCu$_{2}$Si$_{2}$ and CeCu$_{6-x}$Au$_{x}$ where such an analysis gave $\beta$=1.5.~\cite{Stockert07:99}  The critical dynamics in CeCu$_{6-x}$Au$_{x}$ was also found to be described with $\alpha$=0.75 (Ref. \onlinecite{Schroder00:407}), also different from YbRh$_{2}$Si$_{2}$.   While anomalous scaling may exist close to the quantum critical point, our data at Q $\sim$ 0 does not require such forms.  Indeed, the scaling exponent we derive indicates that we are in the quantum disordered regime and higher resolution measurements at fields, temperatures, and wavevectors closer to the critical point are required to make contact with the anomalous thermodynamic data.

 \begin{figure}[t]
\includegraphics[width=8.8cm] {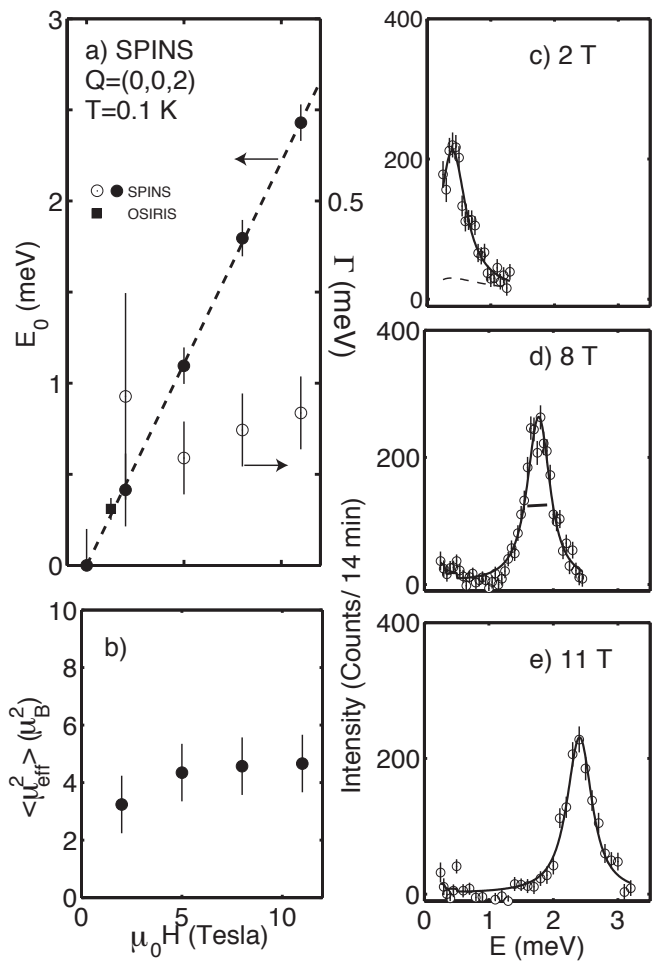}
\caption{\label{B_depend} $(a)$ illustrates the magnetic field dependence of the resonance energy and linewidth.  $(b)$ shows the integrated intensity in absolute units.  Representative constant Q scans are shown in panels $(c)$-$(e)$ at various magnetic fields.}
\end{figure}
  
Next we study the low temperature incommensurate fluctuations as a function of field.  The result is summarized in Fig. \ref{resonance_figure} for an  [1$\overline{1}$0] field where the critical field is $\sim$ 0.06 T.   Fig. \ref{resonance_figure} $(a)$ illustrates that at 5 T a  spin resonance develops with the solid line a fit to an underdamped simple harmonic oscillator plus a relaxational form to describe the low-energy dynamics.  $(b)$ and $(c)$ plot constant energy scans at the resonance energy (1.05 meV) and below at 0.5 meV.  The resonance is found to be  localized in Q space with the fluctuations below the resonance retaining the incommensurate structure measured at zero field.   The solid curve in $(b)$ is a resolution convolved fit to a lorentzian squared resulting in a dynamic correlation length of $\xi$=6 $\pm$ 2 \AA. This indicates that the resonance involves a finite region of correlated spins.  $(d)$ illustrates a contour plot taken on OSIRIS at 1.25 T demonstrating that the resonance peak (even at small magnetic fields) is localized in energy and momentum.  Searches for dispersing spin excitations located in $q$ away from the resonance peak, as measured in ferromagnetic MnSi (Ref. \onlinecite{Ishikawa77:16}) and calcium ruthenate (Ref. \onlinecite{Steffens11:83}) under an applied magnetic field in the paramagnetic state, failed to observe any dispersing modes.  Therefore, the resonance peak measured here is not indicative of field induced spin-waves, like in other itinerant ferromagnets, but rather represents a coherent precession of spin density extending beyond a $4f$ site.

The magnetic field dependence of the resonance peak is displayed in Fig. \ref{B_depend}.  $(a)$ shows that the resonance energy varies as $g\mu_{B}\mu_{0}H$ with $g$=3.8 $\pm$0.2 and remains underdamped at all measured fields with little change in the linewidth.  Background corrected constant Q scans tracking the resonance with field are shown in panels $(c)$-$(e)$.  The integrated weight is also field independent as shown in $(b)$ and matches the calculated spectral weight for intra doublet transitions (or transverse fluctuations) presented in the supplementary information. Therefore all of the Yb$^{3+}$ ions are contributing.  The $g$ factor is comparable to 3.6 obtained from ESR (Ref. \onlinecite{Duque09:79}) and is consistent with the crystal field scheme described in the supplementary information.  Therefore, the ESR signal and the neutron scattering resonance have a common origin and are both associated with the field split ground state doublet of the Yb$^{3+}$ ions.
 
 The presence of both a sharp magnetic field induced resonance and an ESR signal are unique to YbRh$_{2}$Si$_{2}$ amongst heavy fermion materials.  In heavy electron systems, an ESR signal is typically only observed in the presence of local Kondo impurities.  At the phemenological level,  the neutron experimental result is similar to the Haldane spin chain Y$_{2}$BaNi$_{1-x}$Mg$_{x}$O$_{5}$ with Mg impurities.~\cite{Kenzelmann03:90}  In that system, a magnetic field was found to induce a resonance peak which is sharp in energy and momentum representing a staggered magnetization near the edge of a Haldane chain segment.  The finite linewidth in momentum was associated with a dynamic correlation length which measures the extent of the impurity edge state.  Similar arguments have been applied to low energy resonant field induced effects in underdoped cuprates, possibly the result of exciting free spins near charge rich regions.~\cite{Stock09:79}  
 
 A similar physical picture maybe applied to YbRh$_{2}$Si$_{2}$.  A magnetic field induces localized droplets of Yb$^{3+}$ spins which can be resonantly excited through intra doublet transitions.  The localized region of spins are analogous to a Kondo impurity which would give rise to an ESR signal.  The length scale for such a Kondo impurity spin, should be related to $\xi \sim \hbar v_{f}/k_{B}T_{Kondo}=\hbar^{2} k_{f}/mk_{B}T_{Kondo}$.~\cite{Affleck01:86}   Taking T$_{Kondo}$=24 K, $k_{f}$ $\sim$0.5 $a^{*}$, and $\gamma \sim$ 1.5 J/mol K$^{2}$ (Ref. \onlinecite{Trov00:85} at 0.1 K at zero field) yields a lengthscale of $\sim$ 15 \AA.  The result is on the same order as the dynamic correlation length of 6 \AA\ extracted above indicating that $\xi$ maybe set by the Kondo temperature and Fermi velocity.  Because the effect is not due to purely localized Yb$^{3+}$ ions and is not associated with correlated dynamics over long length scales, we refer to this resonance as mesoscopic.  The lack of Fermi surface nesting at high fields, the large Fermi velocity, and the heavy nature of the bands appears to result in the formation of Kondo screened ions.  Such effects appear to be supported by ferromagnetic fluctuations and similar results may exist in other ferromagnetic heavy fermion metals.~\cite{Krellner08:100,Kochelaev09:72,Abrahams08:78}
 
Our experiments indicate that the low temperature zero field dynamics of YbRh$_{2}$Si$_{2}$ is effected by Fermi surface nesting while at high fields and temperatures the response mimics Kondo screened spins above a ferromagnetic transition.    
 
We acknowledge funding from the STFC and the NSF through DMR-0116585 and DMR-0944772.  Work at IQM was supported by DoE, Office of Basic Energy Sciences, Division of Materials Sciences and Engineering under Award DE-FG02-08ER46544. Part of this work was carried out at the Brookhaven National Laboratory which is operated for the US Department of Energy by Brookhaven Science Associates (DE-Ac02-98CH10886) (R.H. and C. P.).  We thank J. Murray and Z. Tesanovic for discussions and R. Down and E. Fitzgerald for cryogenic support.


\end{document}


\title{Supplementary information for ``From incommensurate correlations to  mesoscopic spin resonance in YbRh$_{2}$Si$_{2}$"}

\date{\today}

\begin{abstract}

We present supplementary information regarding the sample preparation and experimental details, crystal field analysis used to provide a benchmark for the neutron scattering cross sections, and further information regarding the scaling analysis presented in the main text. 

\end{abstract}

\maketitle

\subsection{Supplemental material describing experimental setup and samples}

\renewcommand{\thefigure}{S1}
 \begin{figure}[b]
\includegraphics[width=8.5cm] {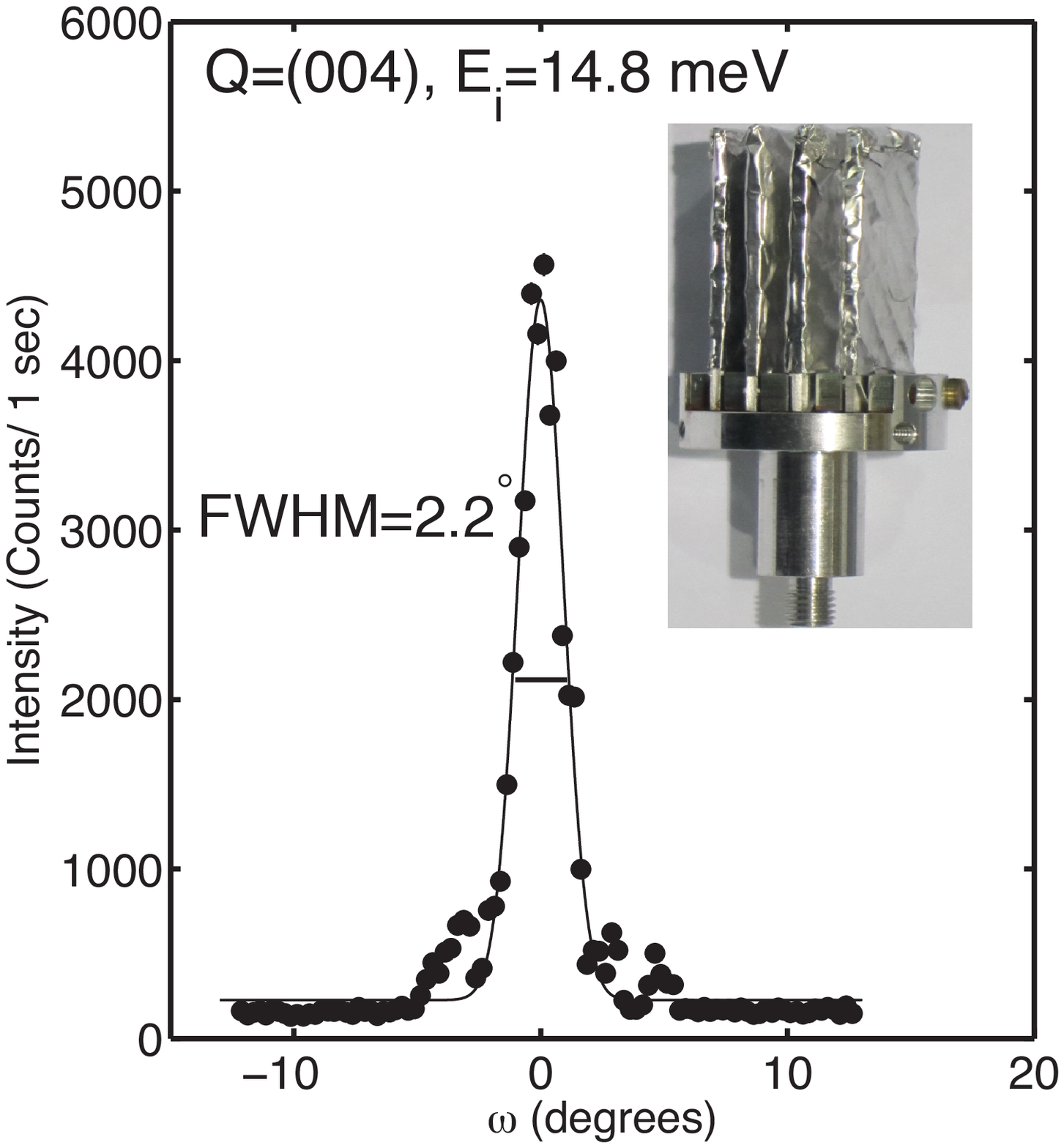} {}
\caption{\label{sample_mount} The rocking curve measured at {\bf{Q}}=(0,0,4) taken on SPINS with E$_{i}$=14.8meV.  The inset illustrates the Aluminium foil covered plate geometry of the sample mount.  The $c$-axis was aligned perpendicular to the plate face. The crystals were grown using a Zinc flux as described in Ref. \onlinecite{Hu07:304}.}
\end{figure}

One of the largest hurdles to obtaining neutron inelastic scattering data on many heavy fermion systems is the possibility to synthesize large single crystals on the order of several grams.  We have overcome this obstacle in YbRh$_{2}$Si$_{2}$ by coaligning $\sim$ 300 samples on aluminium plates using the facets and the plate like geometry of the single crystals.  Each sample was fixed to the plates with Fomblin hydrogen-free oil and further secured by wrapping the plate in aluminum foil.  The resulting rocking curve and sample mount are illustrated in Fig. \ref{sample_mount}.  The rocking curve was measured on SPINS using a configuration of $guide-PG(004)-80-S-80-detector$ at room temperature with E$_{i}$=14.8 meV.

The individual samples were prepared by the same method as described in Ref. \onlinecite{Hu07:304}.  The samples used here were grown from Zinc flux which facilitated the growth of larger samples than produced using indium.  A detailed comparison of the magnetic properties, between samples prepared using both both methods, characterized with ESR is in Ref. \onlinecite{Duque09:79}.  It was shown that, while the residual resistivities of the zinc prepared samples are larger than in materials used in early studies, indium flux typically produces higher resistivity ratios and sharper ESR linewidths.   Heat capacity studies (Ref. \onlinecite{Hu07:304}) did show that the samples prepared using zinc displayed the same divergence as reported for indium prepared samples.  The susceptibility at temperatures above $\sim$ 1 K was also found to be in good agreement between samples prepared using zinc and indium.  

\subsection{Supplemental material describing crystal fields}

\renewcommand{\thefigure}{S2}
 \begin{figure}[t]
\includegraphics[width=8.5cm] {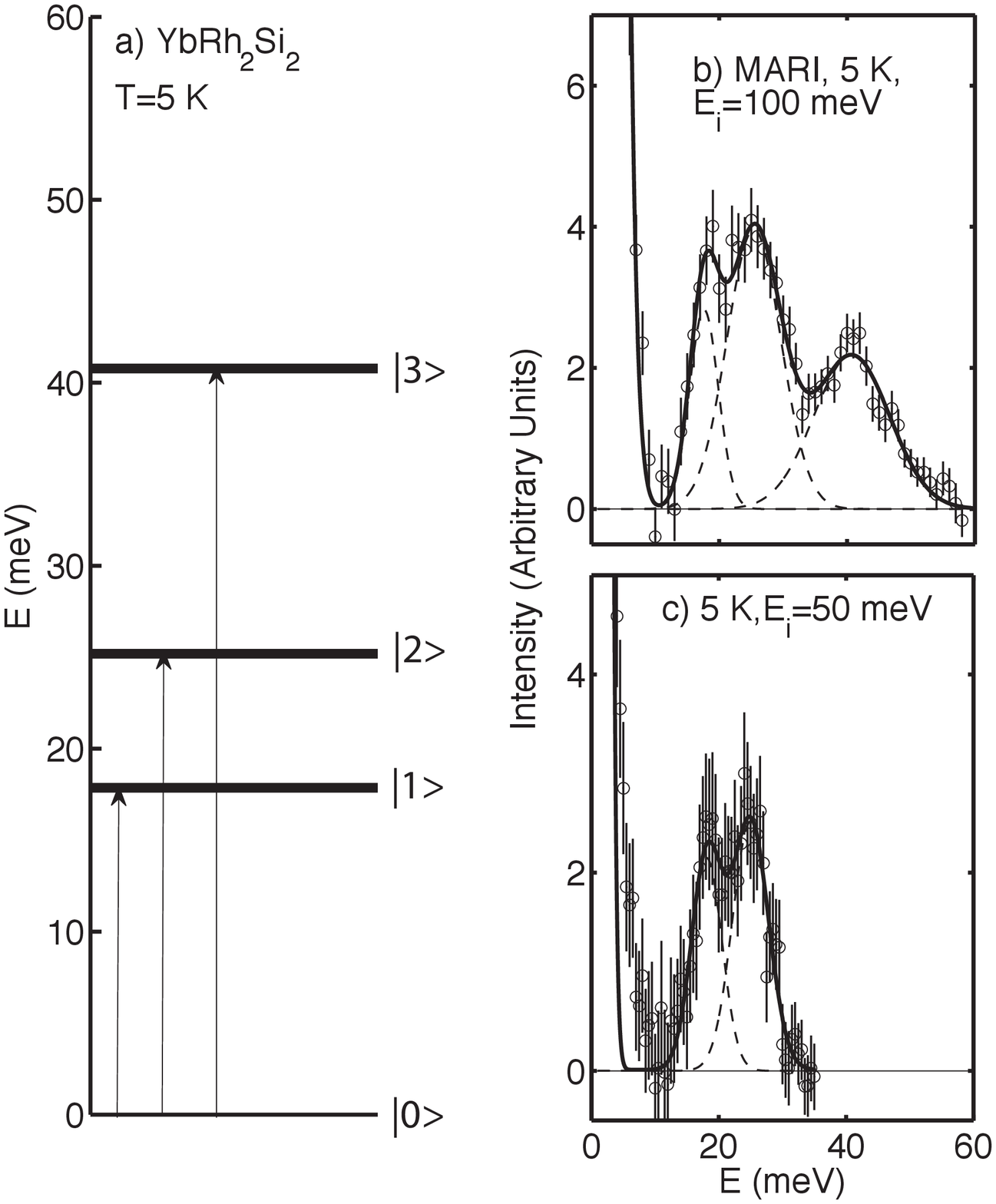} {}
\caption{\label{crystal_fields} The crystal field scheme measured on MARI at 5 K.  $(a)$ is a schematic illustration of the ground state consisting of 4 doublets.  $(b)$ and $(c)$ show constant Q cuts from MARI data at E$_{i}$=100 meV (50 meV), integrating over Q=[1.5, 2.5] \AA$^{-1}$ ([2, 3] \AA$^{-1}$).  The data have been corrected for a background resulting from phonons taken at high momentum transfers and temperature.}
\end{figure}

	In this supplemental section, we present a self consistent description of the crystal field energies, intensities, Lande $g$ factors along with the small ordered moment and the integrated transverse spectral weight reported above.  A crystal field analysis has been discussed in Refs. \onlinecite{Kutuzov11:324,Leushin08:403}, however, for completeness, a similar analysis is shown here in the context of our neutron measurements with the goal of understanding the integrated spectral weight.

	The crystal field excitations were measured on the MARI chopper spectrometer with incident energies of 50 meV and 100 meV.  The powder sample was loaded into a thin annulus to minimize effects due to absorption.  Further scans using E$_{i}$=200 meV found no scattering at energies above 50 meV energy transfer.  To model the background from phonons, we have performed similar scans at large wave vector transfers (integrating over [10, 12] \AA$^{-1}$ and [6.5, 7.5] \AA$^{-1}$ for E$_{i}$=100 and 50 meV respectively) at 150 K where the scattering is dominated by lattice vibrations and the Yb$^{3+}$ form factor ensures minimal magnetic scattering.  These scans were then scaled to the low temperature and low-$Q$ data such that low and high energy transfers were subtracted to zero.  The resulting data in Fig. \ref{crystal_fields} illustrating three distinct crystal field excitations will be the basis for the subsequent analysis.

For Yb$^{3+}$ with $J=7/2$ in a local C$_{4\nu}$ environment, the crystal field scheme should consist of 4 doublets. The corresponding Hamiltonian takes the following form,

\begin{eqnarray}
H_{cef}=B_{2}^{0} O_{2}^{0}+B_{4}^{0} O_{4}^{0}+B_{4}^{4}O_{4}^{4}+B_{6}^{0}O_{6}^{0}+B_{6}^{4}O_{6}^{4}, \nonumber
\end{eqnarray}

\noindent where $O_{n}^{m}$ are the Stevens operators.~\cite{WALTER:1984vt}

\renewcommand{\thefigure}{S3}
 \begin{figure}[t]
\includegraphics[width=9.5cm] {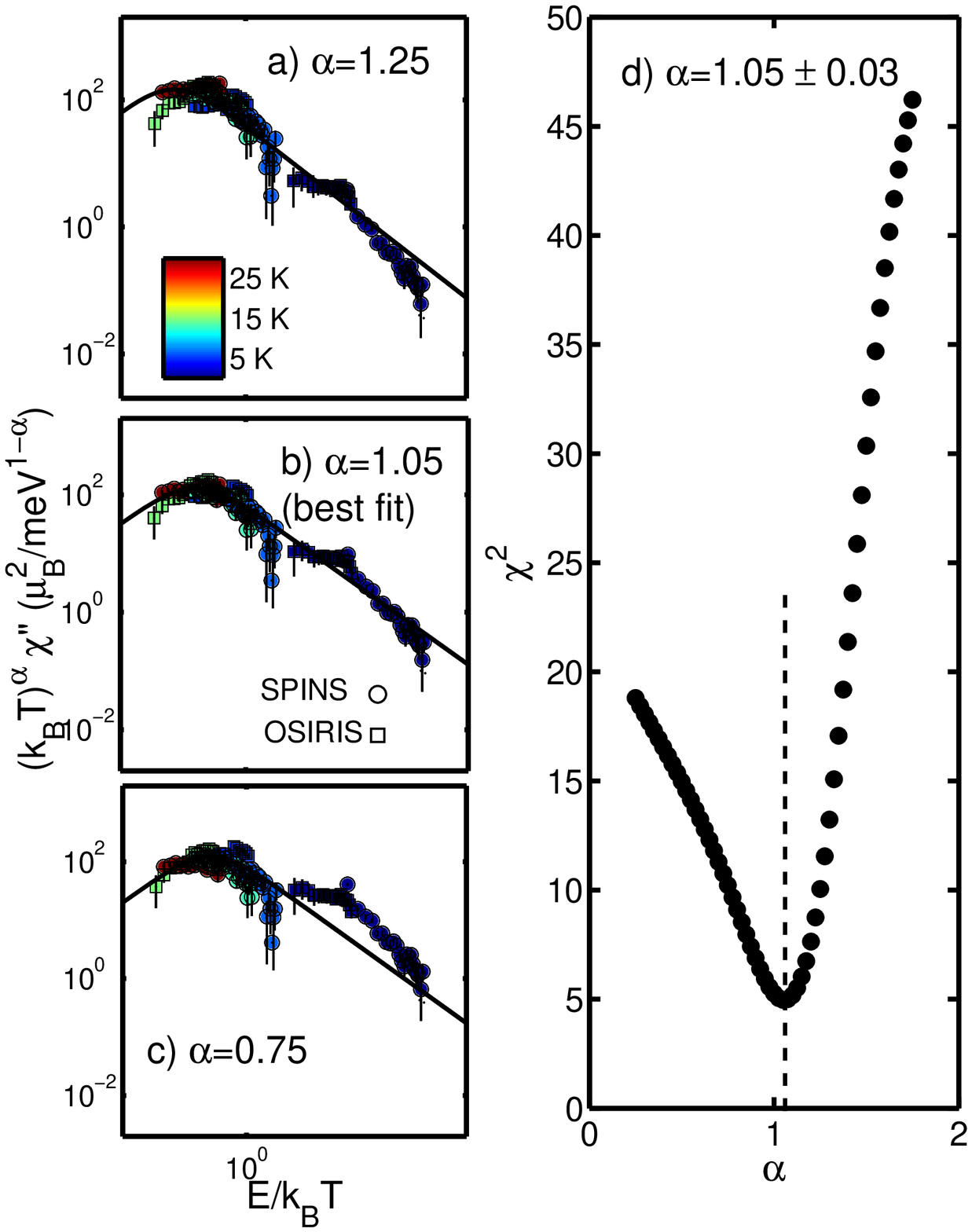} {}
\caption{\label{scaling_error} $a)-c)$ illustrate a series of scaling plots for various fixed values of $\alpha$ over all temperatures and energy range studied.  $d)$ shows a plot of $\chi^{2}$ as a function of $\alpha$ demonstrating that the minimum (denoted by the dashed line) exists at $\alpha$=1.05.  The concavity is a measure of the errorbar.}
\end{figure}

To calculate the experimental values of the Stevens parameters ($B_{n}^{m}$), we have fit the data in Fig. \ref{crystal_fields} to three Gaussians to extract three integrated intensities and three energies.  The crystal field excitations are broader than the resolution of the spectrometer - particularly at higher energy transfers where the resolution on MARI improves over lower energy transfers.  This might be indicative of a finite lifetime for the excitations, or possibly the considerable dispersion noted in recent ARPES studies.~\cite{Vyalikh:2010bk}  For the lower two crystal field levels, we have averaged the results from the E$_{i}$=100 and 50 meV data sets.   An additional data point from neutron scattering is provided by Fig. 4 from which an in-plane  $g$ factor was extracted.  We have also included a $g$ factor parallel to $c$ described in Ref. \onlinecite{Sichel03:91}. This resulted in 8 ``data points" from which the 5 Steven's parameters were fit.  The result is tabulated below.

\begin{table}[h!]
\begin{tabular}{c|c|}
 \hline
$B_{2}^{0}$ & 0.58 $\pm$ 0.10 meV\\
$B_{4}^{0}$ & 0.0122 $\pm$ 0.008 meV\\
$B_{4}^{4}$ & ($\pm$) -0.047 $\pm$ 0.011 meV\\
$B_{6}^{0}$ &  -3.0 e-4$\pm$ 0.8 e-4 meV\\
$B_{6}^{4}$ &   ($\pm$) 0.0102 $\pm$ 0.0016 meV\\
 \hline
\end{tabular}
\label{table_cef}
\end{table}

An analytic analysis found that the energy eigenvalues of the crystal field Hamiltonian only depends on cross terms proportional to the product of $B_{4}^{4}$ and $B_{6}^{4}$. This  introduces a sign ambiguity of both parameters in our fitting analysis and is emphasized in the table by $(\pm)$.   A point charge (Ref. \onlinecite{Hutch06:74}) analysis summing over a cluster of $\sim$ 1000 unit cells shows that $B_{4}^{4}$ has the opposite sign to that of $B_{2}^{0}$ and confirms that $B_{4}^{4}$ and  $B_{4}^{6}$ have opposite signs.  While the applicability of the point charge model to YbRh$_{2}$Si$_{2}$ is debatable owing to strong bonding (Ref. \onlinecite{Hoffmann85:89}), it indicates $B_{4}^{4}$ is negative and $B_{4}^{6}$ is positive.

	Based upon the parameters listed above, the calculated positions of the three doublet excitations are 17.9, 25.2, 40.8 meV and the in-plane $g$ factor is 3.4.  These should be compared with the measured values of 17.9 $\pm$ 0.6, 25.2 $\pm$ 0.6, 40.8 $\pm$ 0.9 meV and the in plane $g$ factor of 3.8 $\pm$ 0.2.  The $g$ factor along $c$ is predicted to be 0.20 consistent with 0.17 $\pm$ 0.07 reported by electron spin resonance measurements.~\cite{Sichel03:91}    The resulting ground state doublet has the following form (for the two solutions listed above),

\begin{eqnarray}
|0_{+}\rangle = (-0.77 \pm 0.05) |{3 \over 2}\rangle + (-0.63 \pm 0.05) |-{5\over 2}\rangle, \nonumber \\
|0_{-}\rangle =(0.63 \pm 0.05) |{5\over 2}\rangle+(0.77 \pm 0.05) |-{3\over 2}\rangle,\nonumber 
\end{eqnarray}

\noindent This results in a low temperature ordered moment of $g|\langle 0|J_{z}| 0\rangle| \mu_{B}=0.010 \pm 0.003 \mu_{B}$  and a transverse cross section (corresponding to intra doublet transitions) of  $ \sum_{i=x,y} |\langle -|J_{i}|+\rangle|^{2} \mu_{B}^2= 5.8 \pm 1.0 \mu_{B}^2$.  The small ordered moment is consistent with $\mu SR$ measurements suggesting values of $\sim$ 10$^{-2}$-10$^{-4}$ $\mu_{B}$.~\cite{Ishida03:68}  The transverse cross section agrees within error with the value reported above to reside in the field induced resonance peak.  This corroborates our claim that the field induced resonance results from transverse spin fluctuations within the lowest energy doublet.

\subsection{Supplemental material describing scaling analysis}

In this section we provide supplementary details of the scaling analysis described above.  The scaling analysis is presented over a broad range in temperature (0.1 K-30 K) with the data from SPINS derived from constant $Q$ scans at {\bf{Q}}=(0,0,2) and the data from OSIRIS integrated over {\bf{Q}}=($\pm$0.1,$\pm$0.1,2$\pm$0.25).  The experiments were performed at temperatures above the magnetic ordering transition expected at $\sim$ 0.05 K.  No magnetic Bragg peak or diverging correlation length was observed over this temperature range and therefore it is not clear whether the experiment is in the critical regime to observe anomalous critical dynamics.  We do find the temperature dependence of the magnetic dynamics to be well represented by a relaxational form  $\chi''({\bf{Q}},\omega)=\chi'_{\bf{Q}}\omega\Gamma(T)/(\Gamma(T)^2+\omega^2)$, where $\Gamma(T)$ is the relaxation rate and $\chi'_{\bf{Q}}$ is the susceptibility.  For an itinerant ferromagnetic at T$\textgreater$T$_{Curie}$, $\Gamma \propto 1/\chi'_{\bf{q}}$ and $\chi'_{\bf{q}}$ has a Curie form ($\propto 1/T$).  The fact that all of the data collapse onto this curve implies that this scaling form is correct.

Deviations from this conventional scaling form have been reported in other heavy fermion systems (namely CeCu$_{2}$Si$_{2}$ and CeCu$_{6-x}$Au$_{x}$) and therefore it is important to quantify how well our data fits to the conventional scaling formalism described above.  Such deviations are quantified in the exponent $\alpha$ and Fig. \ref{scaling_error} illustrates how the errorbar on $\alpha$ was derived.  Panels $a-c)$ show how the data scale as a function $E/k_{B}T$ for several different exponents $\alpha$.  Panel $c)$ demonstrates that while the high temperature data are very well described with $\alpha$=0.75, the low temperature data, closest to the critical point, deviate significantly from this scaling form.  A plot of the normalized $\chi^{2}$, a measure of the quality of the scaling analysis, demonstrates that $\alpha$=1.05 provides the best fit to the data.  From the concavity, we derive an errorbar $\pm$0.03.  Therefore, based on this analysis we do not observe anomalous scaling in the magnetic dynamics in YbRh$_{2}$Si$_{2}$ and find the temperature dependence to be described by relaxational dynamics expected in an itinerant ferromagnet.